%% file: ms.tex
\documentclass[apj]{emulateapj}
\usepackage{graphics}
\usepackage{enumitem}
\usepackage{amsmath}
\usepackage{xcolor}
\usepackage{soul}

\newcommand{\herschel}{\textit{Herschel }}

\def\simlt{\lower.5ex\hbox{$\; \buildrel < \over \sim \;$}}
\def\simgt{\lower.5ex\hbox{$\; \buildrel > \over \sim \;$}}
\def\kms{km~s$^{-1}$}
\def\msun{{$\rm M_\odot$}}
\def\nion{n_{\rm ion}}

\newcommand{\ctwoline}{[\ion{C}{2}]~157.7~$\mu$m} 
\newcommand{\ooneline}{[\ion{O}{1}]~63.2~$\mu$m} 
\newcommand{\othreeline}{[\ion{O}{3}]~88.4~$\mu$m}

\begin{document}

\shortauthors{TEMIM ET AL.}

\shorttitle{Innermost Ejecta in Kes 75}

\title{Probing the Innermost Ejecta Layers in SNR Kes 75: Implications for the Supernova Progenitor}

\author{Tea Temim\altaffilmark{1}, Patrick Slane\altaffilmark{2}, Tuguldur Sukhbold\altaffilmark{3,4}, Bon-Chul Koo\altaffilmark{5}, John C. Raymond\altaffilmark{2},  Joseph D. Gelfand\altaffilmark{6}}

\altaffiltext{1}{Space Telescope Science Institute, 3700 San Martin Drive, Baltimore, MD 21218, USA, ttemim@stsci.edu}
\altaffiltext{2}{Harvard-Smithsonian Center for Astrophysics, 60 Garden Street, Cambridge, MA 02138, USA}
\altaffiltext{3}{Department of Astronomy, The Ohio State University, Columbus, OH 43210, USA}
\altaffiltext{4}{NASA Hubble Fellow}
\altaffiltext{5}{Department of Physics and Astronomy, Seoul National University, Seoul 151-747, Republic of Korea}
\altaffiltext{6}{NYU Abu Dhabi, United Arab Emirates ; NYU Center for Cosmology and Particle Physics, New York, NY 10003, USA}

\begin{abstract}

Supernova remnants (SNRs) that contain pulsar wind nebulae (PWNe) are characterized by distinct evolutionary stages. In very young systems, the PWN drives a shock into the innermost supernova (SN) material, giving rise to low-excitation lines and an infrared (IR) continuum from heated dust grains. These observational signatures make it possible to cleanly measure the properties of the deepest SN ejecta layers that can, in turn, provide constraints on the SN progenitor. We present \textit{Herschel} Space Observatory far-IR observations of the PWN in the Galactic SNR Kes 75, containing the youngest known pulsar that exhibited magnetar-like activity. We detect highly-broadened oxygen and carbon line emission that arises from the SN ejecta encountered by the PWN. We also detect a small amount (a few times $10^{-3}$ \msun) of shock-heated dust that spatially coincides with the ejecta material and was likely formed in the SN explosion. We use hydrodynamical models to simulate the evolution of Kes 75 and find that the PWN has so far swept up 0.05--0.1 \msun\ of SN ejecta. Using explosion and nucleosynthesis models for different progenitor masses in combination with shock models, we compare the predicted far-IR emission with the observed line intensities and find that lower mass and explosion energy SN progenitors with mildly mixed ejecta profiles and comparable abundance fractions of carbon and oxygen are favored over higher mass ones. We conclude that Kes 75 likely resulted from an 8--12 \msun\ progenitor, providing further evidence that lower energy explosions of such progenitors can give rise to magnetars.

\end{abstract}

\section{Introduction} \label{intro}

Connecting supernova remnants (SNRs) to the supernova (SN) explosion sub-types and progenitors that produced them remains a challenge. A number of methods exist for differentiating SNRs resulting from Type Ia explosions from those resulting from core collapse, including the presence of a central compact object or a pulsar, iron and oxygen abundances, SNR morphology and asymmetry \citep{lopez09}, Fe-K line luminosities and energy centroids \citep{yamaguchi14, patnaude15}, and light echoes \citep{krause08a, krause08b, rest08}. Rough estimates of the progenitor masses of core-collapse SNRs have been made using the observed Fe/Si ratio that is sensitive to the CO core mass \citep{katsuda18} and by surveys of stellar populations in the vicinity of SNRs \citep[e.g.][]{kim13,auchettl18}. However, pinning down additional progenitor and explosion properties, such as the SN subtype, has been possible for only a handful of very young SNRs whose abundances and ejecta morphologies can be studied in great detail or for which light echoes have been observed, as in the case of Cas~A \citep[e.g.][]{milisavljevic17}.

\begin{figure*}
\center
\includegraphics[width=6.3in]{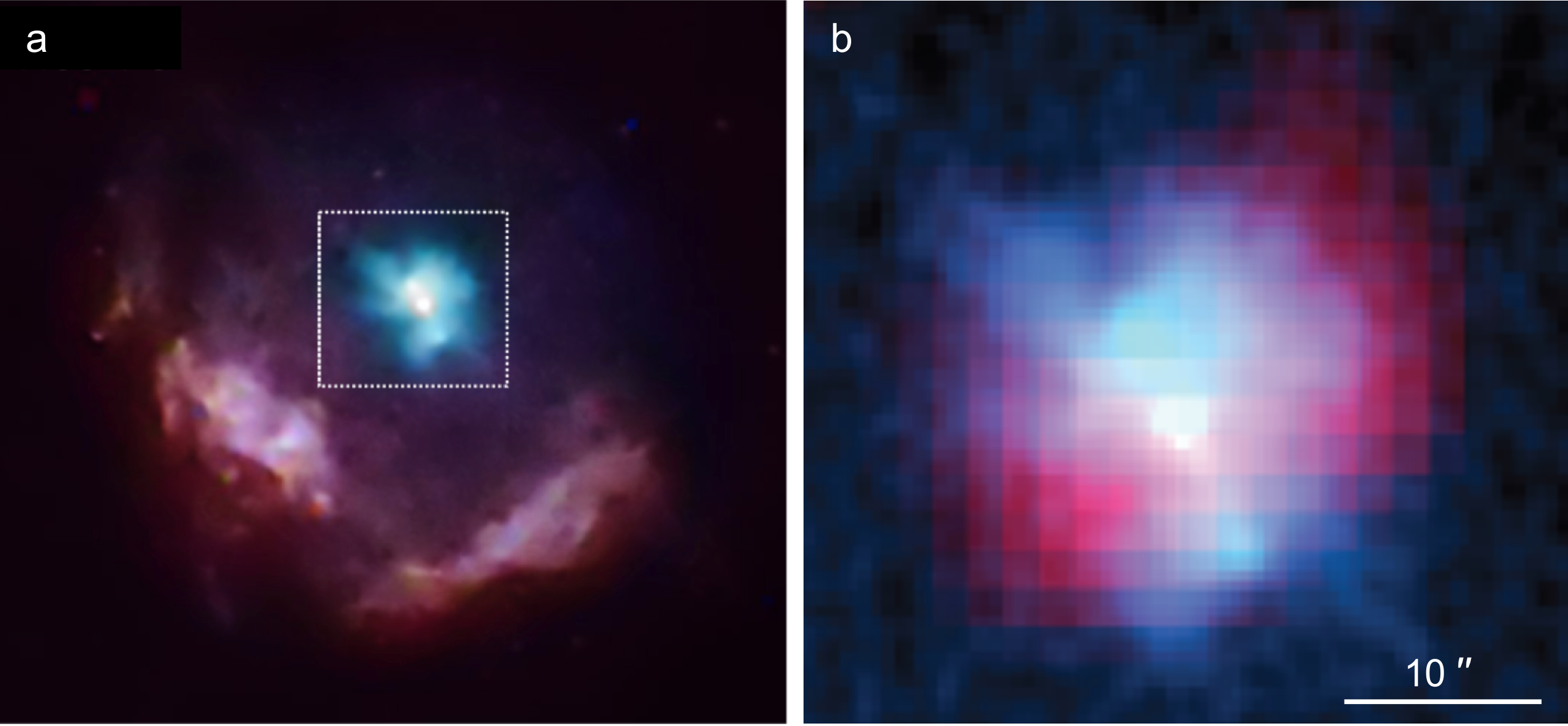}
\caption{\label{fig1} a. \textit{Chandra} X-ray image of Kes 75 with the soft thermal emission shown in violet and hard non-thermal emission from the PWN shown in blue \citep[NASA/CXC/GSFC/][]{gavriil08}. The region in the white rectangle is zoomed in in panel (b). b. Zoomed in view of the PWN in Kes 75 with the non-thermal \textit{Chandra} X-ray emission shown in blue and the \textit{Herschel} 70~\micron\ emission shown in red. The infrared emission seems to be distributed in two lobes along the equatorial region of the PWN, perpendicular to the pulsar's jets (also see Figure~\ref{fig2}.}
\end{figure*}

SNRs of core-collapse explosions that contain young pulsar wind nebulae (PWNe) offer a unique opportunity to probe the innermost SN ejecta layers that can shed light on the progenitor and explosion properties. Early in the SNR's evolution, the PWN drives a shock into the freely-expanding SN ejecta, making them potentially observable at infrared, optical, and X-ray wavelengths through shock-heating and photoionization. Detailed studies of photoionized ejecta filaments in the Crab Nebula provided constraints on the total ejecta mass and explosion energy, identifying the Crab as most likely originating from a sub-energetic, electron capture SN \citep{yang15,jerkstrand15}. 
Studies of the surrounding ejecta and/or dust emission have also been carried out for young PWNe in 3C~58 \citep{slane04,fesen08}, B0540--69.3 \citep{williams08}, G54.1+0.3 \citep{temim10,temim17}, and G21.5--0.9 \citep{zajczyk12, guest19}, providing some constraints on their progenitor masses and explosion types.

Kes 75 (G29.7--0.3) is a Galactic, composite SNR that contains the youngest known pulsar (PSR J1846--0258) and PWN inside a partial thermal shell \citep{becker83}. The distance estimates to the SNR have varied significantly, ranging from 5 to 21~kpc \citep{caswell75,milne79, becker84,leahy08,su09}. The most recent analysis based on \ion{H}{1} observations places the pulsar at a distance of 5.8~$\pm$~0.5~kpc \citep{verbiest12}. \citet{reynolds18} recently used multi-epoch X-ray observations to measure the expansion rate of the PWN in Kes 75. Assuming the latest distance estimate of 5.8~kpc, they found a PWN expansion velocity of $\sim$1000~$\rm km\:s^{-1}$ and calculated the true age of the pulsar to be 480~$\pm$~50~yr.

The X-ray-discovered pulsar \citep{gotthelf00} has a current period of 328~ms and a braking index of n~=~2.19 \citep{archibald15}. 
It has an exceptionally high spindown luminosity of $8\times10^{36}\:\rm erg\:s^{-1}$ and a magnetic field of $5\times10^{13}\:\rm G$, typical for magnetic-field-decay-powered magnetars \citep{gotthelf00}. 
The pulsar indeed underwent magnetar-like X-ray bursts in 2006 \citep{gavriil08}, making the properties of its progenitor of particular interest. 
Kes~75 was hypothesized to have resulted from a Type~Ib/c SN explosion \citep{chevalier05}, primarily due to the high ejecta velocities and low densities implied by the original distance estimate of $\sim$~19~kpc. On the other hand, \citet{reynolds18} propose that Kes~75 likely resulted from a more typical Type~IIP explosion, with the PWN expanding into an asymmetric nickel bubble. While there is evidence that energetic Ib/c SNe may be powered by magnetars \citep{thompson04,metzger11,mazzali14,margalit18,milisavljevic18}, there is also evidence that the birth of magnetars may not require particularly energetic or unusual SNe \citep{vink06,borkowski17,sukhbold17}. Constraining the Kes~75 progenitor and explosion properties would therefore shed light on the SN type/magnetar connection. 

In this paper, we present a detection of the innermost SN ejecta in Kes~75 with the \textit{Herschel} Space Observatory, and use the observations in combination with hydrodynamical (HD), explosion nucleosynthesis, and shock models to investigate the SN progenitor properties.

\section{Observations and Data Analysis} \label{obsv}

\begin{figure*}
\center
\includegraphics[width=6.5in]{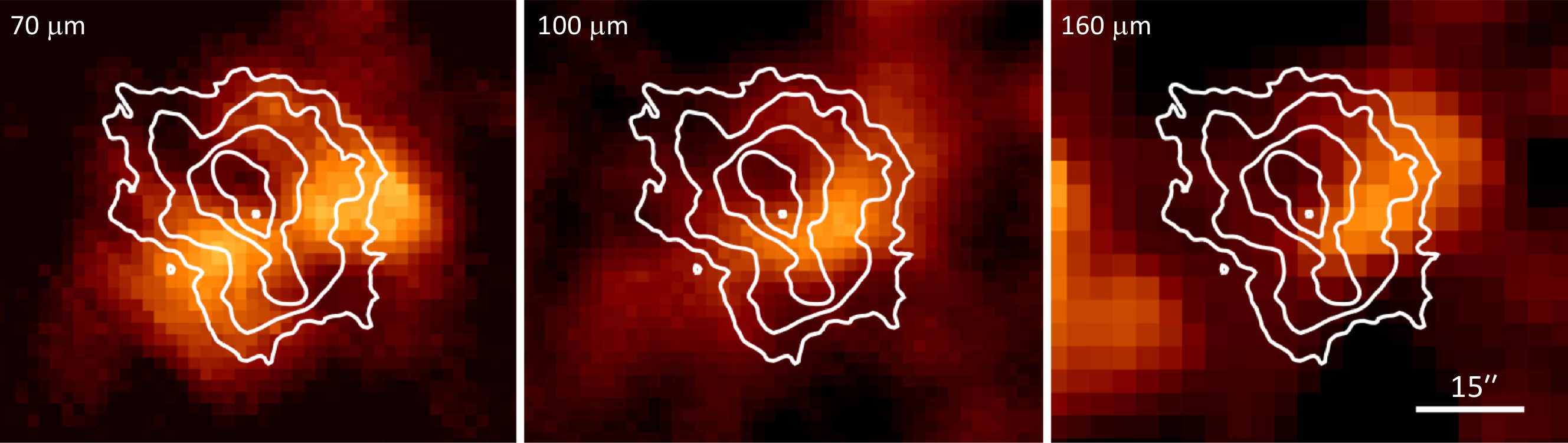}
\caption{\label{fig2}\textit{Herschel} PACS images of the PWN region in Kes 75 at 70, 100, and 160~\micron. The white contours represent the \textit{Chandra} X-ray emission from the PWN. The corresponding measured flux densities are listed in Table~\ref{pacsflux}.}
\end{figure*}

\herschel imaging of Kes 75 was obtained with the Photodetector Array Camera \citep[PACS;][]{poglitsch10} at 70, 100, 160 \micron. The observations were performed on 2012 Oct 9 using a total of 12 scans in the ``scan map'' model and a cross-scan step size of 20\arcsec\ (proposal ID:  OT1\_ttemim\_2, observations IDs: 1342231919-1342231922). The observations for each band were separated into two astronomical observation requests (AORs) with orientation angles of $45^{\circ}$ and $135^{\circ}$. The imaging observations were processed and reduced with the \textit{Herschel} Interactive Processing Environment \citep[HIPE;][]{ott10} version 15.0.1 and images produced using the \textit{MADmap} software \citep{cantalupo09}. The resulting PACS images are in units of Jy/pixel, with pixel scales
of 1.6, 1.6, and 3.2\arcsec/pixel, and full width at half maxima (FWHM) of 6, 8, and 12\arcsec, for the 70, 100, and 160 \micron\ images, respectively. These images are shown in Figure~\ref{fig1}b and Figure~\ref{fig2}.

The flux densities were measured from the PACS images using a circular aperture with a radius of 25\arcsec\ centered on the PWN. The background level was measured from an annular region with the same center and inner and outer radii of 25\arcsec\ and 45\arcsec, respectively. The resulting background-subtracted flux densities are listed in Table~\ref{pacsflux}. The calibration uncertainties for the PACS images were assumed to be 10\%. However, the uncertainties on the measured flux densities are dominated by the uncertainties in the background emission that varies spatially.

\herschel spectra were obtained with the PACS Integral Field Unit (IFU) Spectrometer \citep{poglitsch10} on 2012 Oct 21 using the ``range spectroscopy" mode that covered the \ooneline\ and 145.5~\micron, \othreeline, and \ctwoline\ lines. The IFU has 5 $\times$ 5 spaxels, each measuring 9\farcs4 on a side, and giving a total field of view of 47\arcsec\ $\times$ 47\arcsec\ that covered the entire PWN in Kes~75. An additional pointing with the same parameters was acquired on an off-source background. A total of eight IFU cubes corresponding to the four observed spectral lines and their respective background measurements were analyzed using HIPE version 15.0.1. The spectral edges of each cube were trimmed and the baseline fitted with a 2-degree polynomial across the spectral range that excludes the observed emission lines. The baseline was then subtracted from the cubes. While the baseline-subtracted cubes from the background pointing contained a narrow line component, three of the cubes centered on the PWN contained a broad line component likely arising from SN ejecta in addition to the narrow-line emission. Broad-line emission was detected from \ooneline, \othreeline, and \ctwoline\ lines, but not the [\ion{O}{1}]~145.5~\micron\ line. The lack of detection from the [\ion{O}{1}]~145.5~\micron\ line is not unusual considering that the ratio of the [\ion{O}{1}]~145.5 to 63.2~\micron\ line intensities is expected to be $<$~0.1 \citep{dare97,delzanna15}.

\begin{figure*}
\center
\includegraphics[width=6.5in]{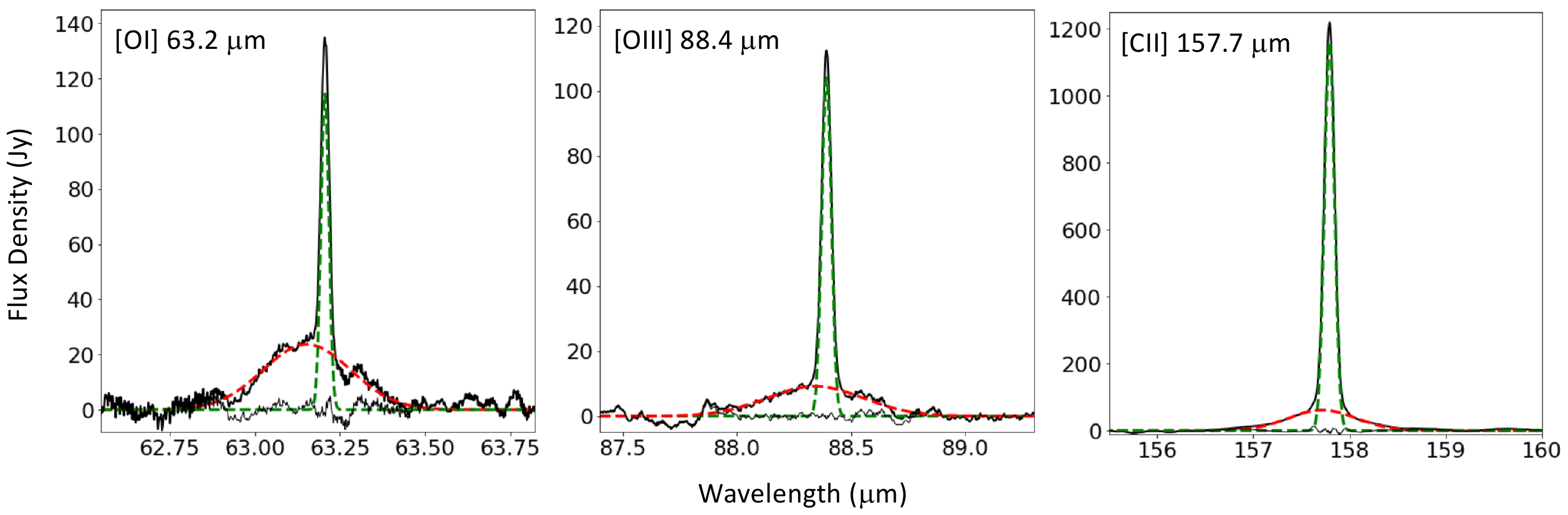} 
\caption{\label{fig3}\textit{Herschel} PACS spectra covering the [\ion{O}{1}] 63.2~\micron, [\ion{O}{3}] 88.4~\micron, and [\ion{C}{2}] 157.7~\micron\ lines are shown in black. The spectra were extracted from the entire field of view shown in Figure~\ref{fig4}. Each of the spectra were fitted with a two-component emission line model, a narrow component representing line emission from the background (green) and a broad component likely arising from the SN ejecta that surrounds the PWN (red). The best-fit parameters for each component are listed in Table~\ref{linetable}.}
\end{figure*}

\begin{figure*}
\center
\includegraphics[width=6.3in]{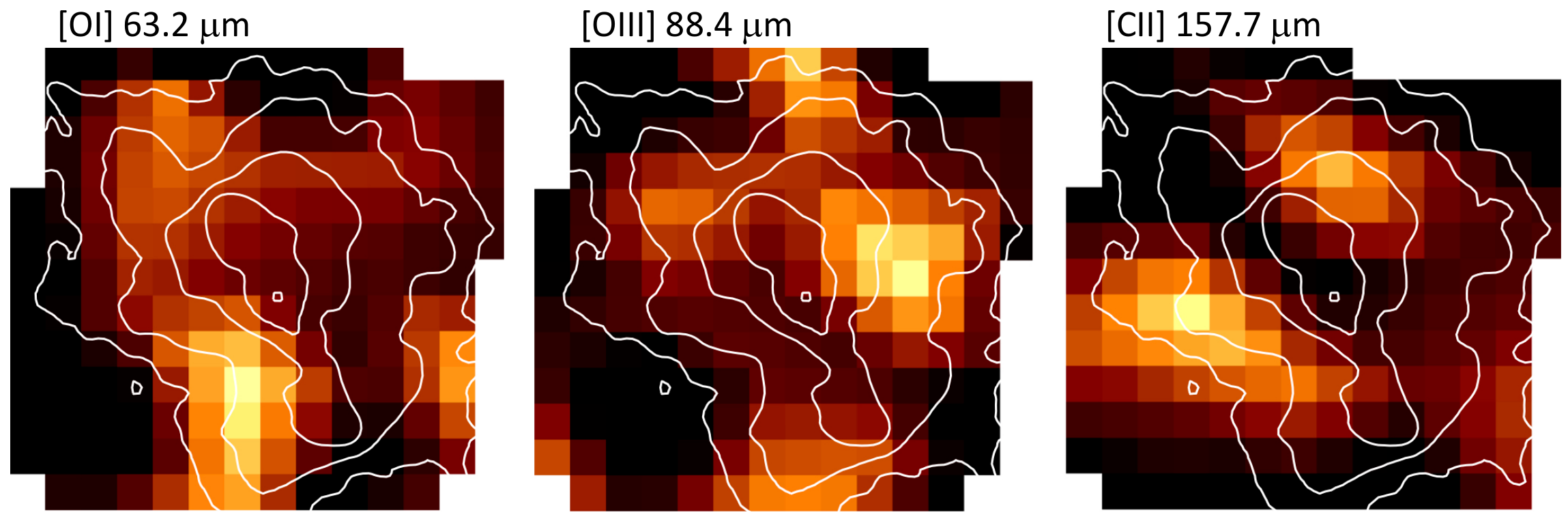} 
\caption{\label{fig4}Emission line maps integrated over the broad spectral line components of [\ion{O}{1}], [\ion{O}{3}], and [\ion{C}{2}] that arise from SN ejecta (red components in Figure~\ref{fig3}) and excluding the narrow-line emission from the background (green components in Figure~\ref{fig3}). White contours represent the \textit{Chandra} X-ray emission from the PWN. All three lines show distinctive peaks within the contours of the PWN.}
\end{figure*}

The final, spatially integrated spectra were produced by summing the spectra from individual spaxels across the entire field of view covered by the IFU. They are shown in Figure~\ref{fig3}. Since the background in the vicinity of Kes 75 is spatially variable, we simultaneously fitted the narrow and broad line components in the PWN spectra instead of directly subtracting the background spectrum. The best-fits are shown in Figure~\ref{fig3} and their parameters listed in Table~\ref{linetable}. We also produced emission line maps that show the spatial distribution of the ejecta emission by integrating the spectra across the broad-line component only, excluding the spectral region across the narrow background line. These maps are shown in Figure~\ref{fig4} with the X-ray contours from the PWN overlaid. 

We also calculated flux densities that each of the broadened ejecta lines contribute to the PACS photometric bands by convolving the best fit spectral model of each ejecta line with the transmission curves for the PACS filters. These values are listed in Table~\ref{pacsflux}. The contribution of line emission ranges from 2--7\% in the 70 and 100 \micron\ photometric bands, and to more than 30\% for the 160~\micron\ band. The estimate does not include narrow line emission arising from the background since this emission has been accounted for by the background subtraction estimated from the annular region.

\input{tab1.tex}


\input{tab2.tex}

\section{Multi-wavelength Morphology} \label{morph}

The \textit{Chandra} X-ray image of Kes~75 is shown in Figure~\ref{fig1}a. There is partial thermal shell with a radius of $\sim$ 1\farcm5 (2.5 pc at a distance of 5.8 kpc) that is prominent in the southeast half of the SNR and completely absent in the northwest. The non-thermal emission from the PWN is evident in the center of the SNR. Radio emission from Kes 75 closely resembles the X-ray morphology and is characterized by a synchrotron spectral index of 0.7 in the shell and 0.25 in the PWN \citep{becker76,helfand03}. The X-ray thermal emission from the shell is well-described by a thermal plasma with a temperature of $\sim$~1.5~keV and solar abundances, consistent with a circumstellar origin \citep{temim09}. \textit{Spitzer} observations of Kes 75 revealed infrared (IR) emission at 24~\micron\ that spatially coincides with the X-ray emission in the shell \citep{morton07}. The IR emission is dominated by $\sim 4\times10^{-3}\:\rm M_{\odot}$ of warm dust that is collisionally heated by the X-ray emitting gas \citep{temim09}.

A zoomed-in image of the X-ray PWN is shown in Figure~\ref{fig1}b. The PWN is elongated along the northeast/southwest axis due to the jets extending in these directions. Its approximate size is 25\arcsec\ and 35\arcsec\ (corresponding to a radius of 0.35~pc and 0.5~pc at a distance of 5.8 kpc) along the short and long axis, respectively. While the \textit{Spitzer} observations of Kes~75 revealed no obvious mid-IR  emission associated with the PWN, the \textit{Herschel} PACS image at 70~\micron\ clearly shows that IR emission surrounds the X-ray PWN. The IR emission appears to be concentrated in two lobes in the equatorial region, perpendicular to the axis of the jets, as can been seen in Figures~\ref{fig1}b and the left panel of Figure~\ref{fig2}. Figure~\ref{fig2} also includes the \textit{Herschel} PACS 100 and 160~\micron\ images that show a distinct peak of emission within the PWN. Based on the morphology alone, the association between the PWN and the IR emission at 100 and 160~\micron\ is less evident than for the 70~\micron\ emission. However, since the emission at 100 and 160~\micron\ peaks in the equatorial plane where the 70~\micron\ emission is concentrated, the association with Kes~75 is likely.

\section{Emission from Supernova Ejecta} \label{ejecta}

The emission lines of oxygen and carbon detected in the \textit{Herschel} spectra have FWHM values in the 1270--1570 $\rm km\:s^{-1}$ range and spatial distribution concentrated in peaks within the PWN contours. This evidence implies that the line emission arises from the innermost SN ejecta that are being encountered by the expanding PWN. The corresponding expansion velocities ($v_{exp}$) for the 63.18  \micron\ [\ion{O}{1}], 88.36 \micron\ [\ion{O}{3}], and 157.74 \micron\ [\ion{C}{2}] lines are 635~$\pm$~50, 775~$\pm$~60, and 785~$\pm$~25~$\rm km\:s^{-1}$, respectively (see Table~\ref{linetable}). 
If the line emission does not arise from a uniform shell, but arises predominantly from ejecta material that has a low tangential velocity, our measured expansion velocity would be a lower limit on the true velocity, which could be as high as $\beta v_{exp}$, where the correction factor $\beta$ is between 1 and 2.

The PWN expansion velocity measured by \citet{reynolds18} is $v_{pwn}$~$\approx$~1000~$\rm km\:s^{-1}$. For a radiative shock, we may expect the measured ejecta velocities to reflect the PWN expansion velocity since most of the emission likely arises from material at the contact discontinuity, where the ejecta density is the highest. Our measured [\ion{O}{3}] and [\ion{C}{2}] line velocities are somewhat lower than the PWN velocity measured by \citet{reynolds18}, unless $\beta$ ranges between 1.3 and 1.6

The measured velocity of the 63.18 \micron\ [\ion{O}{1}] line is slightly lower than for the [\ion{O}{3}] and [\ion{C}{2}] lines, but consistent within the uncertainties. The \ion{O}{1} emission may arise from unshocked ejecta material, in which case its velocity would represent the pre-shock, free-expansion velocity of the SN ejecta. The shock velocity is then given by $v_{shock}=v_{pwn}-v_{ejecta}$. Assuming that the 88.36 \micron\ [\ion{O}{3}] and 157.74 \micron\ [\ion{C}{2}] line velocities are at the PWN velocity,  $v_{shock}$~=~150~$\rm km\:s^{-1}$.
If we take the PWN velocity to be 1000~$\rm km\:s^{-1}$, the correction factor $\beta$ would be 1.28, in which case the ejecta free-expansion velocity of $v_{ejecta}$~=~1.28~$\times$~635~$\rm km\:s^{-1}$~=~810~$\rm km\:s^{-1}$ and the shock velocity $v_{shock}$~=~190~$\rm km\:s^{-1}$.
While the values for the shock velocity are not unreasonable, it is more likely that the emission from neutral oxygen originates from shocked ejecta material that has cooled, in which case its measured expansion velocity would not reflect the free-expansion velocity of the ejecta, but the velocity of the PWN. The relatively lower \ion{O}{1} velocity could then be explained by a different spatial distribution of the \ion{O}{1}-emitting material.

In fact, the spatial distributions of the detected ejecta lines do not appear to be a uniform shell, as can be seen in Figure~\ref{fig4}. While the emission from the three lines is detected across the entire PWN, it is concentrated in distinct brightness peaks that do not coincide spatially.
Possible explanations for the differences in the spatial distribution include an asymmetric distribution of the ejecta that leads to varying compositions in different regions or clumps, a spatial variation in the pre-shock density, or spatially varying line ratios caused by an asymmetric expansion of the PWN that would lead to different shock velocities around the PWN perimeter or different ejecta layers encountered in different regions. For example, \citet{reynolds18} propose that an asymmetry in the explosion may have caused a nickel bubble that is not centered on the expansion center of the PWN. This would not only explain the observed asymmetry in the PWN itself, but would imply that different parts of the PWN expand into different ejecta densities and drive shocks with different velocities. Differences in the spatial distribution may also reflect different levels of photoionization caused by the spatially varying brightness of the PWN.

\section{Emission from Dust} \label{dust}

\begin{figure}
\center
\includegraphics[width=3.3in]{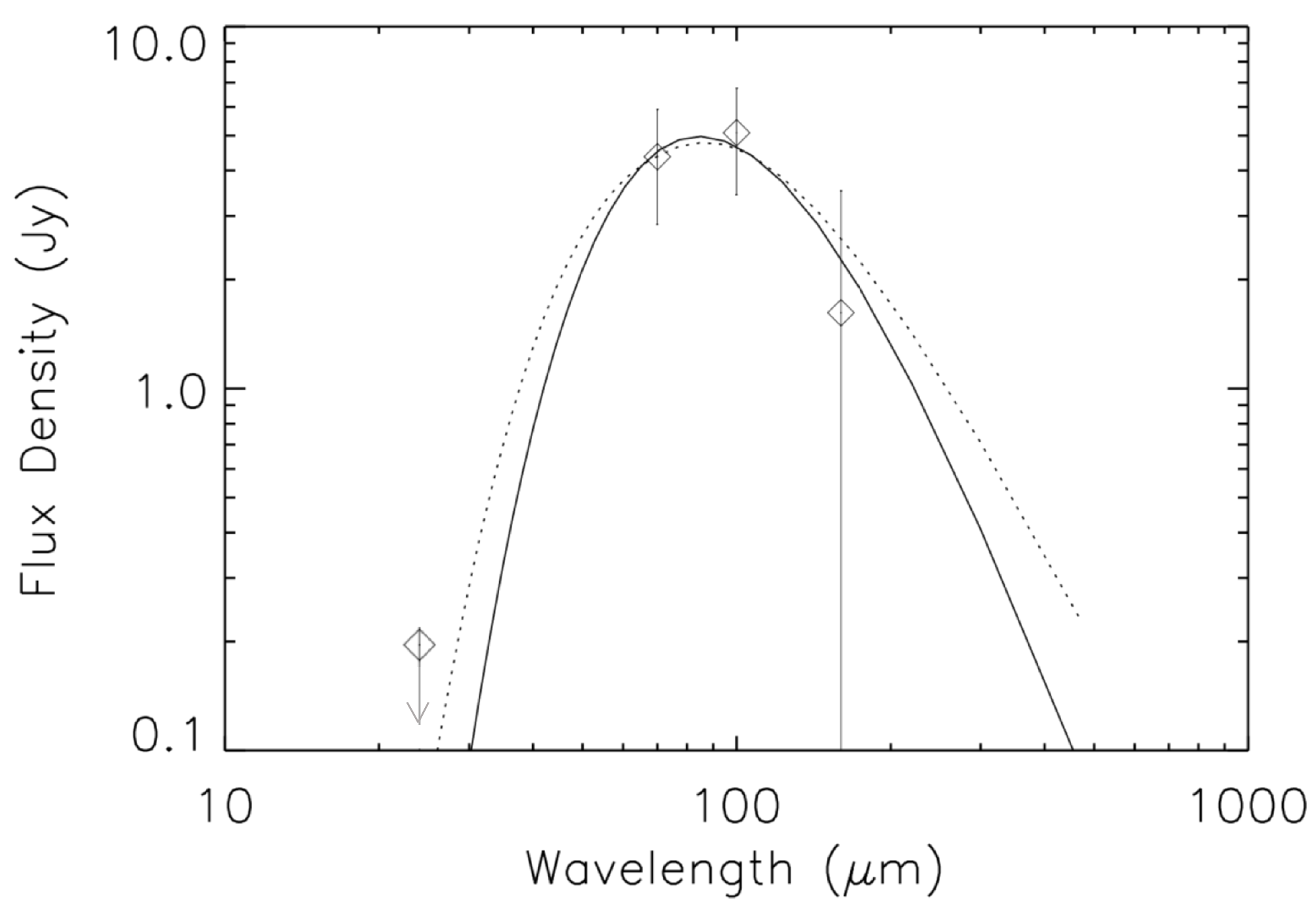}
\caption{\label{fig5}Single-temperature dust models fitted to the\textit{Herschel} PACS SED of the emission associated with the PWN in Kes 75 (flux densities in Table~\ref{pacsflux}). The solid and dotted lines represent the best-fit models for silicate \citep{weingartner01} and amorphous carbon \citep{rouleau91} grain compositions, respectively. The corresponding best-fit dust temperatures and masses are listed in Table~\ref{dusttab}.}
\end{figure}

In a recent study, \citet{omand19} explored the effects of PWNe on dust formation. They found that PWNe produced by pulsars with periods of $\sim$~1~ms and relatively low magnetic fields can delay dust formation and result in a significantly smaller grain sizes, but accelerate dust formation when their magnetic fields are high (B~$10^{15}$~G) due to more effective adiabatic cooling. The pulsar in Kes~75 has a period that is significantly longer, so the PWN's effect on dust formation would likely be negligible. Nevertheless, since any dust present in the ejecta of Kes~75 would have to be newly formed SN dust, characterizing its mass and properties is of significant interest.

The observed flux densities measured from the \textit{Herschel} 70, 100, and 160~\micron\ images and the corresponding contributions from ejecta line emission are listed in Table~\ref{pacsflux}. The emission in the images is dominated by continuum emission, with the ejecta lines contributing 2\%, 7\%, and 33\% to the PACS 70, 100, and 160~\micron\ bands, respectively. In order to estimate the quantity of dust that would be required to produce the continuum emission observed with \textit{Herschel}, we fitted the spectral energy distribution (SED) from the continuum emission (flux densities in Table~\ref{pacsflux} minus the line contribution) with single-temperature carbon \citep{rouleau91} and silicate \citep{weingartner01} grain models. While the \textit{Spitzer} 24~\micron\ image of Kes 75 showed no obvious emission that spatially correlates with the PWN itself \citep{temim12a}, we estimated an upper limit on any PWN-associated emission at 24~\micron\ to be 0.2~Jy. This upper limit was also included in the fit. The SED and the best-fit dust models are shown in Figure~\ref{fig5} and the best-fit parameters for the dust masses and temperatures listed in Table~\ref{dusttab}. For emission dominated by silicate grains, assuming 0.01~\micron\ grains, the best-fit dust temperature and mass are 33~$\pm$~5~K and 0.044~$\pm$~0.041~$\rm M_{\odot}$. For carbon grains, these values are 45~$\pm$~7~K and 0.009~$\pm$~0.006~$\rm M_{\odot}$. Assuming these grain compositions, the dust mass could therefore range anywhere from 0.003 to 0.085~$\rm M_{\odot}$. 

Based on the spatial morphology of the continuum emission, the emitting dust is likely associated with the PWN and originates from ejecta-condensed dust. The emission is concentrated in the equatorial region, perpendicular to the pulsar's jet axis. This is reminiscent of the dust emission in Crab Nebula which is concentrated in the equatorial ejecta filaments. In the case of Kes~75 the spatial distribution of the continuum emission may either reflect the spatial distribution of the dust or a spatial variation in the dust temperature, where the cooler dust emitting at the \textit{Herschel} wavelengths is concentrated in the equatorial region. Since the radius and expansion velocity of the PWN are smaller along this region, the shock velocity is lower and any shock-heated dust may therefore have a lower temperature. In order to determine if the dust is indeed shock heated by the PWN, we calculated the temperature of the same carbon and silicate grain compositions that would result from radiative heating of dust by the PWN. Assuming that the grains are located 0.35~pc from the PWN (the smallest radius to the edge of the PWN of 12.5\arcsec\ at a distance of 5.8~kpc) and that the heating source is the broadband spectrum of the PWN from \citet{gelfand14}, we find that the dust would be heated to a maximum temperature of $\sim$~20~K, even for the very small grains. Since the best-fit dust temperatures to the observed emission are considerably higher than this, we conclude that the dust is most likely shock-heated. As will be discussed in the following sections, the PWN has so far swept-up only between 0.045 and 0.1~$\rm M_{\odot}$ of SN ejecta, so the SN-formed dust is likely to be on the low end of the estimated dust mass distribution, on the order of few times $10^{-3}$~\msun. This would still lead to a significant dust-to-gas mass ratio of at least 0.03. 

\input{tab3.tex}


\section{Kes 75 Progenitor} \label{prog}

In order to understand the implications of the observations presented here, we first need to identify  constraints on the system parameters that will be used to model the dynamical evolution and shock emission that produces the observed IR line fluxes and ratios.
There still exists a significant range in the estimates for the progenitor mass and explosion energy that produced Kes 75. 
Recent theoretical one-zone models for the spectral evolution of the PWN have used observational constraints on the pulsar, PWN, and the SNR to reproduce the observed PWN broadband emission and predict explosion parameters for Kes~75. \citet{bucciantini11} calculated an age range, and a set of explosion energies, ejecta masses, and ambient densities, for an assumed ejecta profile that would give the observed size for the PWN and the SNR shell at a distance of 6~kpc, as well as the forward shock speed of 3700~$\rm km\:s^{-1}$ measured from the Si X-ray line in the SNR shell \citep{helfand03}. Their additional constraint on the ejecta mass to be roughly in the 5--16~$\rm M_{\odot}$ range comes from the assumption that Kes~75 originates from a SN Ib/c explosion. The SN Ib/c explosion has been suggested for Kes 75 mainly due to the fact that the previous distance estimate of 19~kpc implied a very large SN expansion velocity of $\sim$~10,000~$\rm km\:s^{-1}$ and suggested that the SN has been expanding through a region of very low density and was now interacting with a higher density shell \citep{chevalier05}. However, the most recent distance estimate of 5.8~$\pm$~0.5~kpc \citep{verbiest12} alleviates the requirement for such a high expansion velocity. \citet{bucciantini11} find that for a PWN expanding into a  flat ejecta profile, the age varies from 450 to 650 yr and the ejecta mass from 4.8 to 16.4~$M_{\odot}$, respectively. The required explosion energy is $\simeq\:2\:\times\:10^{51}$~erg. Cases with a steeper ejecta profile were permissible for older ages and much lower explosion energies of $\lesssim\:0.3\:\times\:10^{51}$~erg.

\citet{gelfand14} also used a dynamical and spectral one-zone evolution model to estimate the physical properties of the Kes~75 progenitor. Their best-fit parameters imply a slightly sub-energetic, low ejecta mass SN with an explosion energy of $\simeq\:0.8\:\times\:10^{51}$~erg and an ejecta mass of 3.2~$\rm M_{\odot}$.
Using their most recent measurements for the PWN expansion velocity and age (1000~$\rm km\:s^{-1}$ and 480 yr), the radius of PWN of 0.42~pc, and the initial spin-down luminosity of $L_{0}\:\sim\:4\:\times\:10^{37}\: \rm erg\:s^{-1}$, \citet{reynolds18} calculate an upstream ejecta density of $\rho_{ej}\:\sim\:10^{-23}\: \rm gm\:cm^{-3}$ and a swept-up ejecta mass of 0.05~$\rm M_{\odot}$. This result is consistent with a low ejecta mass for the explosion under the assumption of a uniform ejecta distribution. However, \citet{reynolds18} point out that even though the result of the low ejecta density encountered by the PWN is robust,  a uniform ejecta distribution is likely an oversimplification, especially due to the expected nickel bubble effect resulting from radioactive decay of $\rm ^{56}Ni$. The energy from the nickel decay produces a low-density bubble in the innermost region of the SNR that can have a density an order of magnitude lower than in the outer ejecta \citep{chevalier05}. This possibility led \citet{reynolds18} to conclude that, despite the low ejecta density in the innermost region, the total ejecta mass of Kes~75 could be much larger and that the explosion then most likely resulted from an ordinary Type IIp SN. They further point out that for a low-energy and low-ejecta-mass explosion, the innermost density would need to be a couple of orders of magnitude lower than the density inferred for Kes~75.

In summary, the ejecta mass of Kes~75 remains uncertain. If the SNR resulted from a the SN Ib/c, the progenitor may be a single, massive Wolf-Rayet star whose massive winds have removed the bulk of the stellar envelope, or a lower-mass star whose envelope has been stripped away by a binary companion \citep{smith14, dessart15, yoon17}. Kes~75 also could have resulted from a typical Type IIp SN if the PWN is currently expanding into a nickel bubble, or even a Type IIb SN if the progenitor's envelope was stripped by a binary companion.

\section{Modeling}

The goal of the following sections is to use the information from the line emission detected from the innermost ejecta around the PWN, in combination with HD, explosion, nucleosynthesis, and shock models, to provide insight into the properties of the Kes~75 progenitor. We explore the expected abundances and line emission from both low and higher mass progenitors and compare them to the observed line emission in Kes~75. We use the HD models to estimate the mass of the ejecta that has so far been swept-up by the PWN and the current radial distance of the shock for each of the progenitor cases. We then use explosion and nucleosynthesis models to obtain abundance ratios at these radial distances and predict the IR line emission using shock models. We find that the observations are consistent with a low mass progenitor and a mildly mixed ejecta profile. Below, we describe the details of the modeling that led to this conclusion.

\subsection{Hydrodynamical Models}\label{hydro}

\input{tab4.tex}


The details of the interaction of a PWN with the innermost ejecta in an SNR depend on the spin-down properties of the pulsar as well as the SN explosion energy and ejecta mass, and the density of the medium into which the SNR expands. In the simple case of free expansion for both the PWN and the SNR, the PWN radius expands as \citep{chevalier92}

\begin{equation}
R_{\rm PWN} \approx 0.14 \dot{E}_{0,38}^{1/5} E_{51}^{3/10} M_{\rm ej}^{-1/2} t_{100}^{6/5} {\rm\ pc}
\end{equation}
where $\dot{E}_{0,38}$ is the initial pulsar spin-down power in units
of $10^{38}{\rm\ erg\ s}^{-1}$, $E_{51}$ is the explosion energy in units
of $10^{51}$~erg, $M_{\rm ej}$ is the ejecta mass in solar masses,
and $t_{100}$ is the age in units of 100 yr. The associated swept-up
ejecta mass is
\begin{equation}
M_{\rm sw} \approx 4.4 \times 10^{-3} \left(\frac{\dot{E}_{0,38}t_{100}}{E_{51}} 
\right)^{3/5} M_{\rm ej}.
\end{equation}
For typical values of the pulsar and SN parameters, the PWN thus probes only the innermost ejecta regions in the first several hundreds of years.

\begin{figure}
\center
\vspace{4mm}
\includegraphics[width=3.4in]{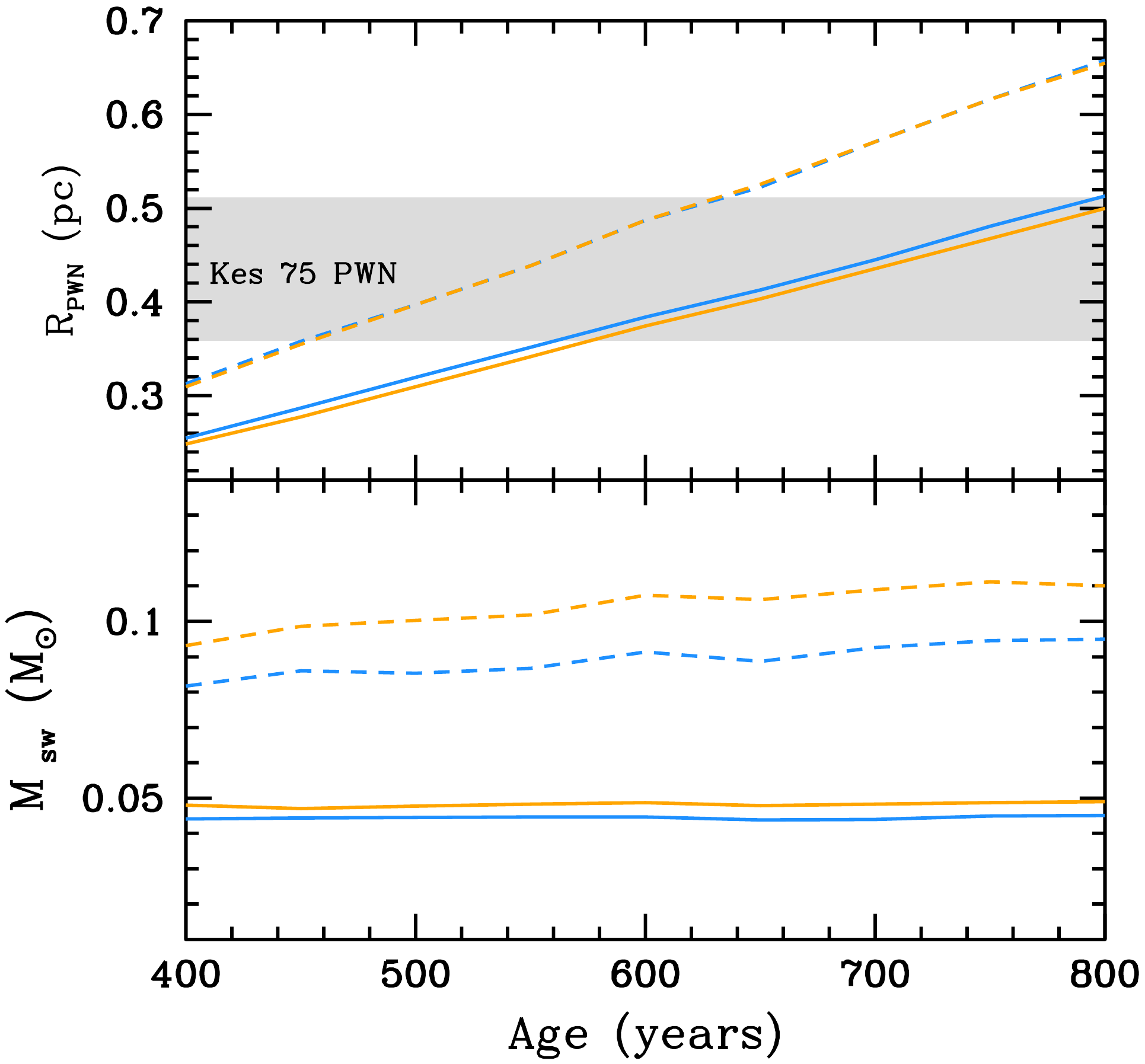}
\caption{\label{mswept} Results from the HD simulations based on two sets of model parameters listed in Table~\ref{hdinput}; the \citet{bucciantini11} model (blue curves) and the 10~\msun\ ZAMS progenitor model (orange curves). The dashed curves represent the case in which the early synchrotron losses by the particles in the PWN are completely ignored, while the
solid curves represent the upper limit on the losses, derived assuming magnetic field properties similar to those inferred for the Crab Nebula at early stages \citep{gelfand09}. The top panel of the figure shows the evolution of the PWN radius with time, while the bottom panel shows the evolution of the swept-up SN ejecta mass. The gray band represents the range in the observed radius of the non-spherical PWN, taken to be 0.36--0.51~pc at a distance of 6.0~kpc assumed by \citet{bucciantini11}. For more details, see Section~\ref{hydro}.
}
\end{figure}

Modeling of the broadband emission from the Kes 75 PWN has been carried out by several groups (see Section~\ref{prog}). To investigate the origin of the IR emission described above, we have carried out 2D HD simulations using the parameters from \citet{bucciantini11}, listed in Table~\ref{hdinput}.
A density gradient in the ambient medium was introduced in order to reproduce the shell size and asymmetry observed in Kes~75 (see Figure~\ref{fig1}a). The ambient density varies from $\rm 0.4d_0$ in the northern shell region to $\rm 1.6d_0$ in the southern shell, where $\rm d_0=2.0\times10^{-24}\:g\:cm^{-3}$. The simulations were carried out using the VH-1 hydrodynamics code \citep{blondin01} using a treatment for the PWN contribution as described in \citet{kolb17}. The results are shown as blue curves in Figure~\ref{mswept}, where the dashed curve ignores early synchrotron losses by the particles in the PWN, while the solid curve assumes that the fraction of the spin-down power lost as synchrotron radiation at early stages is similar to that in models for a Crab-like system \citep{gelfand09}. 

The top panel of Figure~\ref{mswept} shows the evolution of the PWN radius for these two extreme cases. The gray band represents the observed radius range for the non-spherical PWN in Kes~75. The PWN age range is 450--620~yr for no synchrotron losses and 580--800~yr for Crab-like losses. The average PWN expansion velocity for the two cases is 850 and 630~$\rm km\:s^{-1}$, respectively, which is consistent with the velocities derived from the observed line widths (see Table~\ref{linetable} and Section~\ref{ejecta}). The model ejecta density is a few times $\rm 10^{-23}\:g\:cm^{-3}$, while the shock velocities as the PWN overtakes the SN ejecta are $\rm < 100 \: km\:s^{-1}$.
The bottom panel of Figure~\ref{mswept} shows the evolution of the mass of the SN ejecta swept-up by the PWN. For the model with parameters from \citet{bucciantini11} (blue curves), the total swept-up ejecta mass range is 0.045--0.085 \msun\ for Crab-like and no synchrotron losses, respectively.
The actual values for Kes~75 are expected to fall between the two curves, most likely closer to the solid curve and the lower swept-up ejecta mass. We note that the non-monotonic nature of the curves results from uncertainties in determining the exact shock position in the simulations. The fact that the swept-up mass appears essentially flat with time in the case where the early synchrotron losses are large (solid lines) is due to a low shock velocity that leads to a low rate of increase for the mass.

For comparison, we carried out a simulation based on a 10~\msun\ zero-age main sequence (ZAMS) mass progenitor model for which $M_{\rm ej}=8.2$
and $E_{51}=0.6$ \citep{sukhbold16}, described in Section~\ref{nucleo}. We adjusted the ambient density to provide the
observed radius of the southern rim of Kes 75, incorporating the same density gradient as used above, but with $\rm d_0=2.5\times10^{-25}\:g\:cm^{-3}$. The parameters for this model are summarized in Table~\ref{hdinput}. We assumed the same pulsar parameters as determined by \citet{bucciantini11}.  The results for the PWN size evolution are shown in the top panel of Figure~\ref{mswept} (orange curves) and are nearly indistinguishable from those derived from the \citet{bucciantini11} models results (blue curves). The PWN age and expansion velocity ranges are therefore the same as in the \citet{bucciantini11} case described above. The swept-up ejecta mass shown in the bottom panel of Figure~\ref{mswept} ranges between 0.05 and 0.1~\msun\ for the Crab-like and no synchrotron losses, respectively.
While a full investigation of all realistic models for Kes 75 is beyond the scope of this study, our results are consistent with those presented
by \citet{reynolds18}, showing that the PWNe in young SNRs are effective probes of the innermost ejecta products produced in their progenitor's explosions. In the next section, we investigate the innermost abundance profiles resulting from the explosion of the progenitors corresponding to the the two cases explored by the HD simulations.

\subsection{Explosion and Nucleosynthesis Models}\label{nucleo}

\input{tab5.tex}


\begin{figure}
\center
\vspace{4mm}
\includegraphics[width=3.4in]{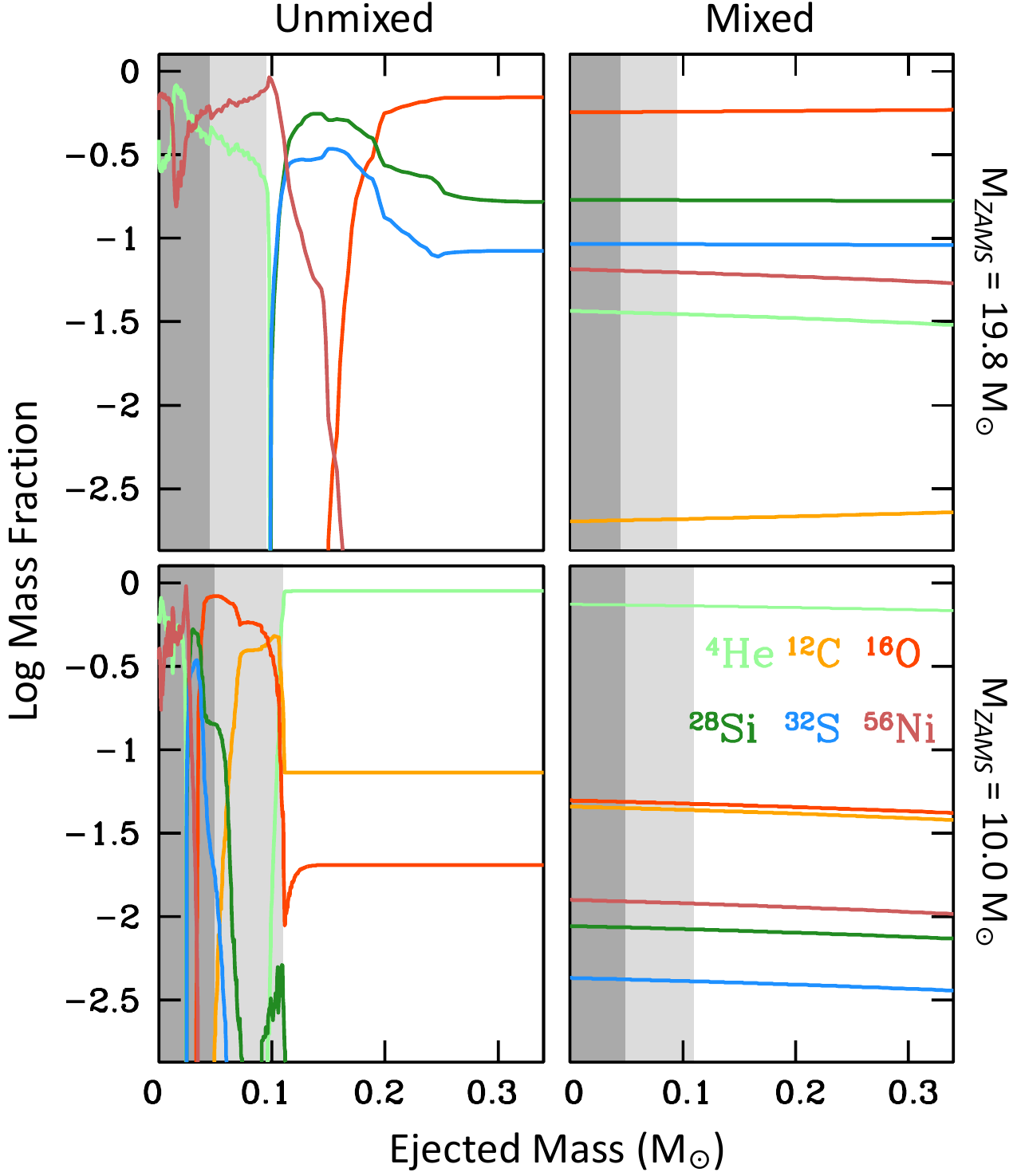}
\caption{\label{explosion}Mass profiles for the unmixed and mixed models of \citet{sukhbold16} for explosions of progenitors with ZAMS masses 19.8~$\rm M_{\odot}$ (top) and 10.0~$\rm M_{\odot}$ (bottom). The x-axes begin outside of the neutron star mass and reflect the material that has been ejected in the explosion. The gray bands represent the amount of mass that has so far been swept up by the PWN in Kes~75, as derived from the HD simulations for each progenitor case. The light gray bands represent the range in the total swept-up mass values assuming the extreme cases of no synchrotron losses and Crab-like synchrotron losses for the PWN, leading to a lower and upper limit on the swept-up ejecta mass of 0.045 and 0.085 \msun\ for the 19.8~\msun\ progenitor and 0.05 and 0.1 \msun\ for the 10.0~\msun\ progenitor, respectively (see Section~\ref{hydro}).
}
\end{figure}

In order to explore the expected abundances in the innermost ejecta layers encountered by the PWN in Kes~75, we employ stellar evolution and explosion models presented in \citet{sukhbold16}, carried out for 200 non-rotating solar metallicity progenitors with initial masses between 9 and 120~$\rm M_{\odot}$. A key novel feature of these models is that, unlike most prior nucleosynthesis surveys, they are free from arbitrary mass cuts and dialed-in explosion energies, and instead, the final fates and the explosion properties, including energy, compact object mass, and nucleosynthesis yields were all uniquely tied to the progenitor core structure. For technical details on these calculations see \citet{sukhbold16} and \citet{ertl16}. The spherically symmetric models include unmixed ejecta mass profiles, as well as mass profiles resulting from extensive artificial mixing applied after all the explosive nucleosynthesis was completed in the ejecta. The simple mixing scheme was crudely tuned to SN~1987A \citep{kasen09}, and attempts to account for the mixing due to turbulent convection and Rayleigh-Taylor instabilities in the ejecta.

We chose two sample successful explosion models that represent the two cases simulated in Section~\ref{hydro}, a progenitor with $M_{\rm ZAMS}=10\ \rm M_{\odot}$ from their ``Z9.6'' engine, and $M_{\rm ZAMS}=19.8\ \rm M_{\odot}$ model from their ``N20'' engine, which has the explosion energy and ejecta mass parameters most similar to the \citet{bucciantini11} case. 
Properties of these progenitors and their explosions are listed in Table~\ref{progenitors}, while the composition of the innermost ejecta both before and after the extensive artificial mixing are illustrated in Figure~\ref{explosion}. The mass profiles on the x-axes of Figure~\ref{explosion} begin outside of the proto-neutron star and only reflect the material that has been ejected. The gray bands represent the amount of mass that has so far been swept up by the PWN in Kes~75, based on the HD results from the previous section. The light gray band represents the uncertainty in the total swept-up ejecta mass derived from the HD models (see Figure~\ref{mswept}), with the lower and upper limit of swept-up ejecta reflecting the range in the assumed synchrotron losses by the particles in the PWN (see Section~\ref{hydro}). In the next section, we use shock models in combination with the ejecta profiles shown in Figure~\ref{explosion} to estimate the intensities of the far-IR lines and compare them to those observed in the Kes~75 PWN.

\subsection{Shock Models}\label{shock}

We have calculated the expected  \ctwoline, \ooneline, and \othreeline\ line intensities from the PWN shock for the 10.0 \msun\ and 19.8 \msun\ progenitor cases described in Section~\ref{nucleo} and shown in Figure~\ref{explosion} in order to compare them to those observed in Kes~75.
Since the unmixed ejecta case for the 19.8 \msun\ progenitor does not contain carbon in the innermost region that has been encountered by the PWN (see the top left panel of Figure~\ref{explosion}), we rule out this model for Kes~75 and exclude it from the shock model calculations.
The shock code that we used is a one-dimensional hydrodynamic code for planar shocks developed by Raymond and his collaborators \citep{raymond79, cox85, blair00}. It is a revised version for the diagnostics of  metal-rich SN ejecta with some updates in atomic parameters.
The code follows a fluid element as it cools after being heated and compressed at the shock front, including time-dependent ionization balance calculation that includes photoionization. The electron and ion fluids are followed separately, with Coulomb collisions transferring energy between them. The ratio of the electron to ion temperature ($T_e/T_i$) at the shock front is taken to be a free parameter considering that the electrons can be heated at the shock front by plasma turbulence or other means \citep{raymond95}. The ionization states of ions entering the shock are taken to be those in photoionization equilibrium under shock radiation, determined by an iterative method. The code does not include thermal conduction which would increase the emission of the lower ionization stages \citep{borkowski90}. A more detailed description of the code can be found in \cite{blair00}.

\input{tab6.tex}


We have run a grid of models for shock speeds $v_s=50$ to 200 \kms\ at 50 \kms\ intervals and preshock mass densities 
$\rho=8.4\times (10^{-23}, 10^{-22}, 10^{-21})$~g cm$^{-3}$. The line intensities also depend on the magnetic field strength $B_0$, but not sensitively. We adopt $B_0=1.0$ $\mu$G. The ratio of electron to ion temperature at the shock front  is fixed to be $T_e/T_i=0.8$ and the computation is stopped when the temperature drops below $\sim 500$ K. For the chemical composition of the preshock gas, we assume three different cases with compositions listed in Table~\ref{abundtab} and plotted in Figure~\ref{explosion}; the mixed ejecta models for the 19.8 and 10.0 \msun\ progenitors and the unmixed model for the 10.0 \msun\ progenitor at the position where the swept-up ejecta mass is $\approx$~0.1~\msun. This particular zone was chosen to be representative of the range where carbon and oxygen mass fractions are comparable (the 0.07-0.1 \msun\ range in the bottom left panel of Figure~\ref{explosion}). 

We note that the mass density of $8.4\times 10^{-23}$ g cm$^{-3}$ corresponds to the ion density ($\nion$) of 15.2 cm$^{-3}$ and 2.9 cm$^{-3}$ for the 10 \msun\ and 19.8 \msun\ progenitor cases, respectively. In the mixed 10~\msun\ progenitor model, hydrogen and helium are the major elements contributing 97\% of the particle number density. The number densities of oxygen and carbon are small (1\%) and comparable. In the unmixed 10~\msun\ model, carbon (53\%) and oxygen (38\%) are the dominant contributors, with the next major elements being neon (6.8\%) and magnesium (1.5\%). In the 19.8~\msun\ progenitor, oxygen is the dominant element contributing 62\% of the number density with helium (16\%) and silicon (11\%) as the next major elements. There is very little carbon (0.3\%) and no hydrogen in the 19.8~\msun\ progenitor.

The results of the shock models are shown in Figure~\ref{shock_models}. The top panels show the predicted intensities of the collisionally excited lines for the three progenitor models and the three different densities, along with the observed line intensities for Kes~75. The observed intensities were derived by dividing the line fluxes from Table~\ref{linetable} by $2\pi A$, where A is the emitting area of the PWN ($\sim$~1018~arcsec$^2$),  in order to match the intensity units of the planer shock model output. The bottom panels of Figure~\ref{shock_models} show the [\ion{C}{2}]/[\ion{O}{1}] line ratios for the three densities (black lines) and the observed ratio for Kes~75 (green line). Considering that the spatial morphologies of the emission line maps differ from each other (see Figure~\ref{fig4}), a single plane parallel shock model only gives very crude average conditions. However, we do note that the [\ion{C}{2}]/[\ion{O}{1}] intensity ratio varies spatially by only a factor of $\sim$~2.

\begin{figure}
\center
\includegraphics[width=3.2in]{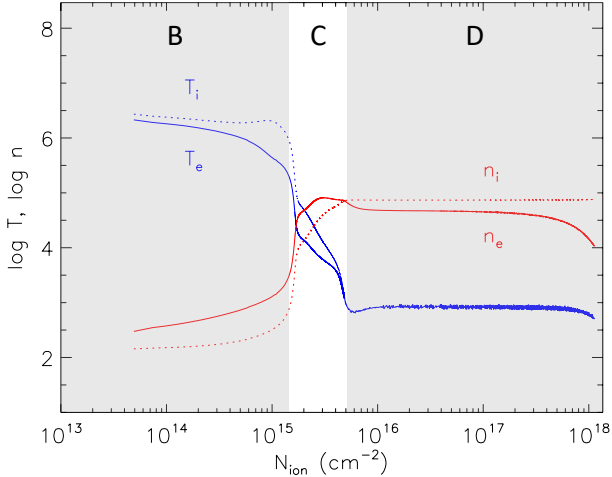}
\caption{\label{td_shock}Temperature and density structure of the 150~\kms shock in the unmixed 10~\msun\ model with $\rho_0=8.4\times 10^{-22}$ g cm$^{-3}$. The density and temperature of the electrons ($n_e, T_e$) and ions $(n_i,T_i)$, as a function of the ion column density swept up by the shock, are plotted as the solid and dashed curves, respectively. The extent of the hot postshock region (B), the thin cooling layer (C), and the photoionization region (D) are indicated by the gray/white bands.}
\end{figure}

\begin{figure*}
\center
\vspace{4mm}
\includegraphics[width=7in]{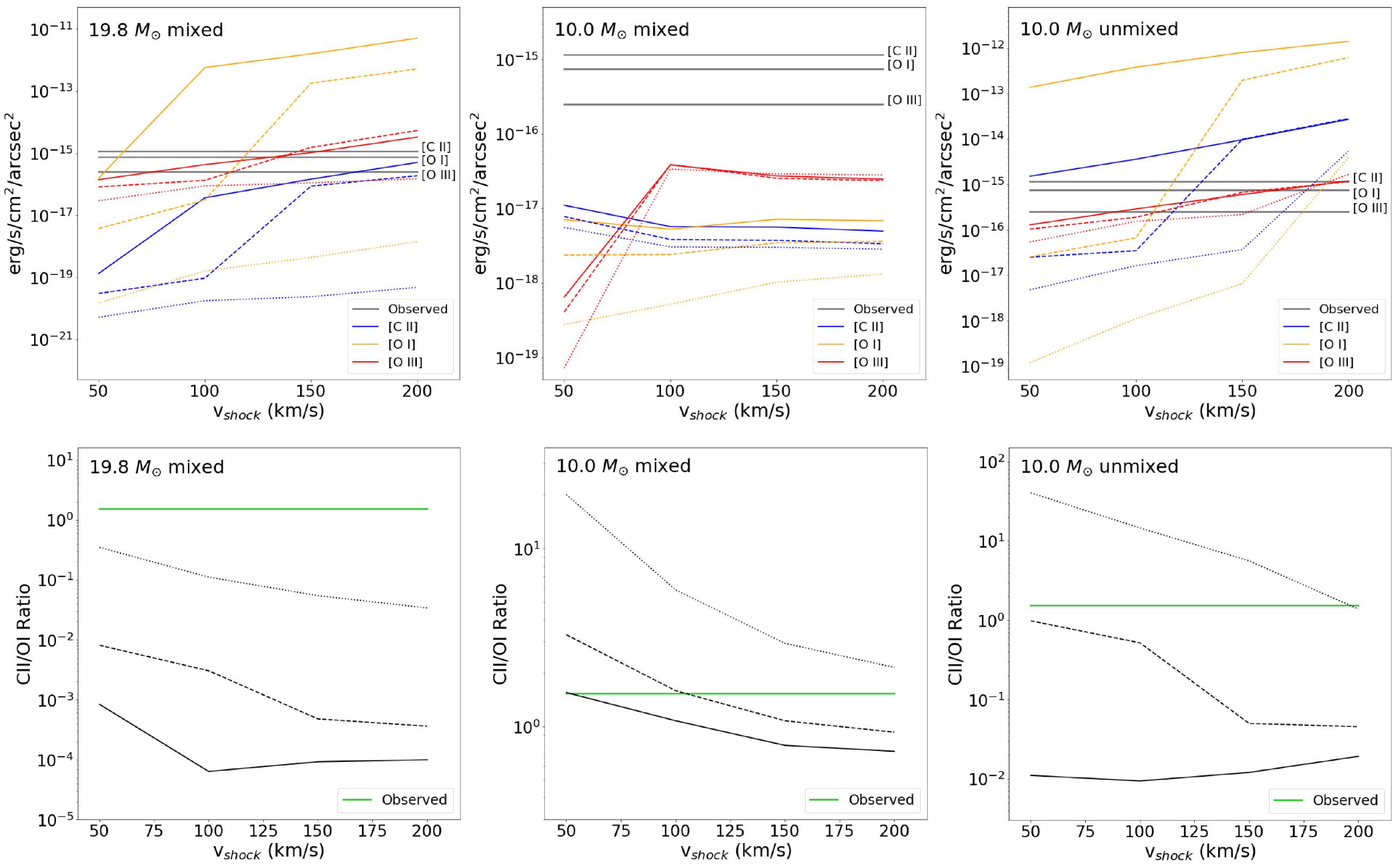}
\caption{\label{shock_models} Top: Modeled line intensities for the \ctwoline\ (blue), \ooneline\ (orange), and \othreeline\ (red), corresponding to the three different cases (see Section~\ref{nucleo}) with compositions listed in Table~\ref{abundtab} and plotted in Figure~\ref{explosion}. The dotted, dashed, and solid lines represent preshock mass densities of $\rho=8.4\times (10^{-23}, 10^{-22}$ and $10^{-21})$~g cm$^{-3}$, respectively. The horizontal gray lines represent the observed line intensities. Bottom: The modeled [\ion{C}{2}]/[\ion{O}{1}] line ratios for the three different densities (black lines) and the observed ratio (green line).}
\end{figure*}

Before we compare the shock model results to the observed line fluxes in Kes 75, it is worthwhile to discuss briefly the shock structure and the emission of the far-IR lines. The structure of the shock in metal-rich plasma has been studied to understand the emission from young, oxygen-rich SNRs such as Cas A or 1E 0102.2$-$7219 in the Large Magellanic Cloud \citep{itoh81, itoh86, dopita84,borkowski90,sutherland95,blair00,docenko10}. Since the 19.8~\msun\ progenitor is oxygen rich,   the results of these studies might be applicable to the current work.  According to these studies, the shock structure can be divided into four regions;  
(A) radiative precursor where the gas is photoionized by shock radiation,
(B) hot post shock region where the electron and ion gases are hot with roughly constant $T_e/T_i$,  
(C) thin cooling layer where the temperature drops abruptly and the ions recombine, and 
(D) extended photoionization region (PIR) where the gas is reionized by EUV radiation from the hot shock region. The temperature and density structure of the shock and the extent of regions B--D are shown in Figure \ref{td_shock}. Region A is upstream of the shock and is not shown in the figure.
The \othreeline\ line originates from the radiative precursor (A) and the cooling layer (C). The radiative precursor has a much lower density than the cooling postshock layer, so with density-sensitive lines one can determine where the emission originates. 
In Cas A ejecta knots, for example, where the shock models with $v_s=140$--200~\kms\ and $n_0=30$--100~cm$^{-3}$ give reasonable fits to the observed line intensities, the contribution from the two regions appears to be comparable \citep{itoh81,sutherland95,docenko10}.
However, the assumption that the photoionization precursor reaches the equilibrium state assumed by the models is questionable. 
Our shock models do not include emission from the photoionization precursor (region A), so the \othreeline\ line flux is likely underestimated by up to a factor of two.

The \ooneline\ and \ctwoline\ lines, on the other hand, mainly originate from the PIR (D), which can be extensive for fast shocks.  
So in Cas A, for example, the shock models with a fully developed PIR overpredict the intensities of the observed \ion{O}{1} recombination lines, and the PIR had to be truncated to match the observed intensities \citep{itoh86,dopita84,borkowski90,blair00}. 
Presumably such an extensive PIR is prevented either by the insufficient columns or by some physical mechanism, e.g., turbulent shredding of the cold gas \citep{blair00}. This is another indication that 1D steady flow models provide only an approximate description of the shock flow, but to the extent that they produce reasonable ratios of ionization fractions for ions of different elements that form at the same temperature, they predict reasonably reliable relative line intensities.

In our shock models for Kes 75, an extensive ($\simgt 10^{18}$~cm$^{-2}$) PIR develops for fast shocks (150--200~$\rm km\:s^{-1}$) in the denser ejecta in all three progenitor cases. The predicted \ooneline\ and \ctwoline\ line intensities from such a fully developed PIR can be orders of magnitude greater than the observed intensities, in which case the PIR would need to be artificially truncated to match the observations. The ratio of these lines, however, is approximately independent of the spatial extent of the PIR and should give meaningful abundance comparisons.  The PIR does not develop for slow/diffuse shocks (50--100~$\rm km\:s^{-1}$) in any of the progenitor cases. In principle, the steady-flow 1D models should include a PIR, but if its temperature is lower than 500 K, it is not captured in our shock models. (Model predictions at those low temperatures would not be reliable without some treatment of molecule formation and cooling.) The predicted intensities of \ooneline\ and \ctwoline\ lines from the slow shock models therefore  only account for the contributions from the cooling layer (C), and as we will see below, tend to underpredict the observed line intensities. This may indicate the presence of a PIR in Kes 75, although it can be partially truncated as in Cas A. If the contribution from a PIR dominates, where O is mostly in O$^0$ and C is mostly C$^+$, the ratio of the \ooneline\ and \ctwoline\ line intensities could be a direct measure of the oxygen and carbon abundances. In the cooling layer, the ions are still in high ionization states in general, so the \ooneline\ line is emitted where O is mostly O$^+$ and therefore the predicted [\ion{C}{2}]/[\ion{O}{1}] line ratio from our shock models will be higher than that from the PIR.

In summary, a successful shock model should be able to reproduce the observed [\ion{C}{2}]/[\ion{O}{1}] line ratio for the higher shock velocities and the absolute \othreeline\ line intensity within a factor of $\sim$~2. The absolute values of the \ooneline\ and  \ctwoline\ line intensities for the higher shock velocities will likely be overestimated by the model, but can be reduced by truncating the extent of the PIR region. 
Since the PIR does not form at the lower shock velocities ($v_{shock}<\:150\:\rm \:km\:s^{-1}$) in our model, for these shock velocities, the absolute \ooneline\ and  \ctwoline\ line intensities can be somewhat underestimated by the model, while the [\ion{C}{2}]/[\ion{O}{1}] line ratio would be somewhat overestimated.

\subsubsection{Comparison with Observations}

Keeping the above discussion in mind, we compare the observed \ctwoline, \ooneline, and \othreeline\ line intensities to the shock model results in Figure~\ref{shock_models} for each of the three progenitor cases.
While the line intensities for the mixed ejecta 19.8~\msun\ model are comparable to the observed line intensities for certain densities and shock velocities, the [\ion{C}{2}]/[\ion{O}{1}] ratio is significantly underpredicted by the model for all densities and across all tested shock velocities. Therefore, regardless of whether the lines originate in the cooling layer or the PIR, the relative carbon and oxygen mass fractions for the 19.8~\msun\ progenitor (Table~\ref{abundtab}) cannot reproduce the observed line ratios (see the bottom left panel of Figure~\ref{shock_models}). 

For the mixed ejecta 10.0~\msun\ model, the [\ion{C}{2}]/[\ion{O}{1}] ratios are in the range of the observed ratio, but the line intensities are significantly underpredicted for all three lines; by an order of magnitude for the \othreeline\ line and two orders of magnitude for the \ctwoline\ and \ooneline\ lines. This is true even for the case of the highest density, which results in the formation of an extensive PIR. While the relative carbon and oxygen mass fractions for this progenitor model are consistent with the observed ratio, the absolute values are too low to reproduce the observed far-IR line intensities.

The results for the mixed ejecta models discussed above indicate that in order to reproduce the observed line intensities and ratios in Kes~75, the ejecta profile requires comparable mass abundances of carbon and oxygen, but with absolute masses that are larger than for the 10.0~\msun\ model. Such properties are offered by the unmixed 10.0~\msun\ model. The lower left panel of Figure~\ref{explosion} shows the mass profiles for the 10.0~\msun\ model before extensive artificial mixing has been applied. The mass profile region between an ejected mass of 0.07 and 0.1~\msun\ has comparable fractions of carbon and oxygen, but absolute fractions that are an order of magnitude larger than for the mixed case shown in the bottom right panel of Figure~\ref{explosion} (also see Table~\ref{abundtab}). We therefore take the fractional masses from the 0.1~\msun\ mass zone to be representative of this region and use them as input to our shock model calculations. We note that this region falls within the light gray band that denotes the range of possible ejecta masses swept up by the PWN, as found from the HD simulations in Section~\ref{hydro}. 

The results of the shock models are shown in the right panels of Figure~\ref{shock_models}. The \othreeline\ line intensity is within the observed range for shock velocities of 150~$\rm km\:s^{-1}$ or below, depending on the chosen density, especially considering that our shock models do not include the emission from the photoionization precursor that would increase the \othreeline\ line intensity by up to a factor of two. The \othreeline\ line intensity is overpredicted for shock velocities higher than 150~$\rm km\:s^{-1}$ for the lowest density and 100~$\rm km\:s^{-1}$ for the higher densities. 

The \ctwoline\ and \ooneline\ line intensities are within the observed range for slow shocks (50-100~$\rm km\:s^{-1}$) for densities between the middle and highest density (solid and dashed lines), but significantly underpredicted for the lowest density (dotted lines). On the other hand, the [\ion{C}{2}]/[\ion{O}{1}] ratio for the higher densities is underpredicted by the model. If a significant PIR is present at the lower shock velocities, the line intensities and ratios would be more consistent with observations for the lowest density, but not for the higher densities.
For the higher shock velocities, at which an extended PIR has developed, the \ctwoline\ and \ooneline\ line intensities are consistent with observations for the lowest density (dotted lines) and overpredicted for the higher densities. While the absolute values of the modeled intensities for these two lines can be varied by either truncating or increasing the extent of the PIR, the [\ion{C}{2}]/[\ion{O}{1}] ratio indicates that the density is likely on the low end, on the order of $10^{-23}-10^{-22} \rm \: g\:cm^{-3}$, consistent with the values from the HD simulations. In summary, the shock model results for the unmixed ejecta 10.0~\msun\ model can reproduce the observed \othreeline\ line intensity for shock velocities $<$~150~$\rm km\: s^{-1}$, as well as the intensities and ratios of the \ctwoline\ and \ooneline\ lines for the lower densities.

\section{Discussion}

\begin{figure}
\center
\vspace{4mm}
\includegraphics[width=3.4in]{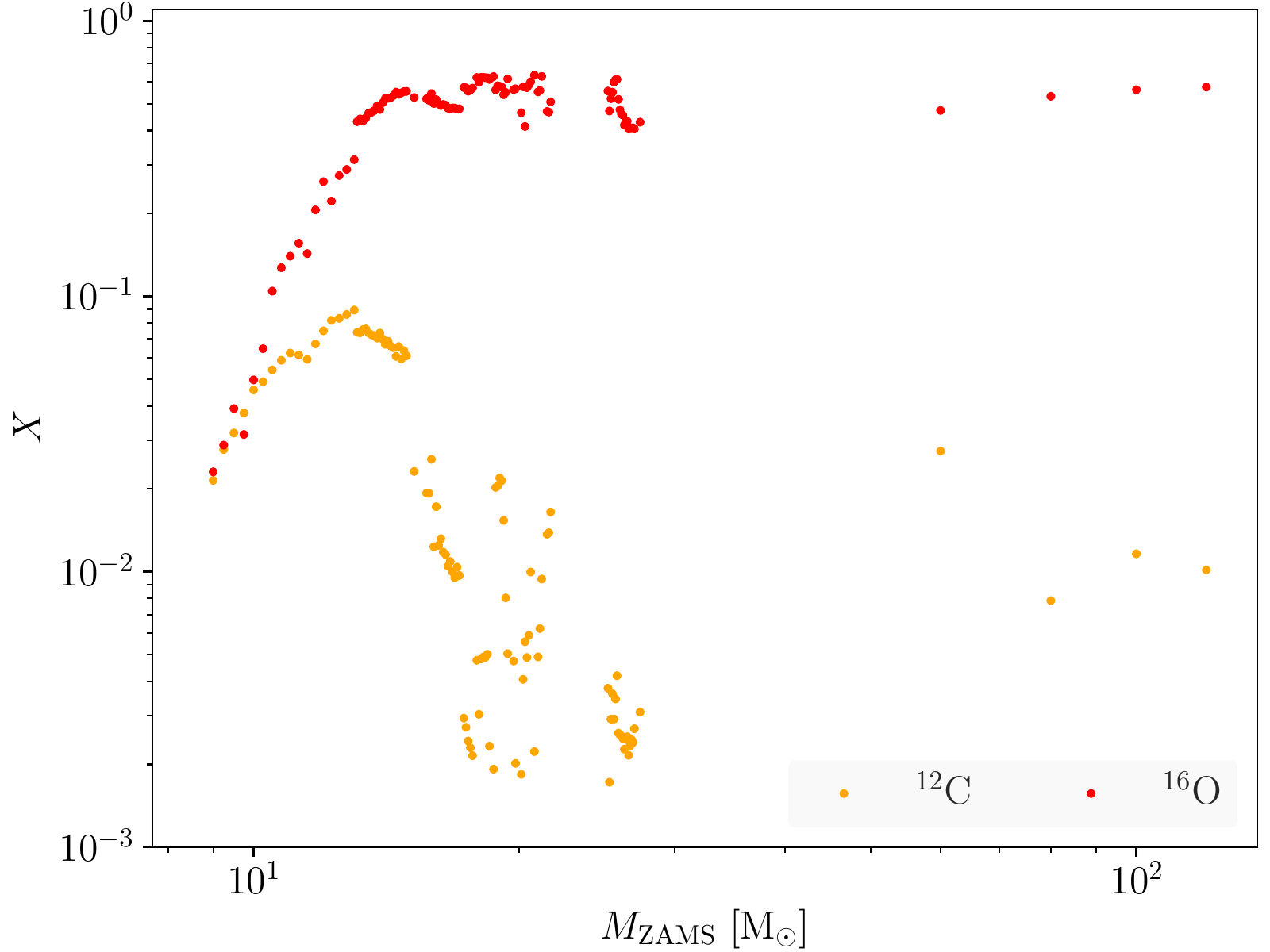}
\caption{\label{fig10}Carbon and oxygen mass fractions for mixed ejecta models of \citet{sukhbold16} as a function of the progenitor ZAMS mass ($M_{ZAMS}$). The orange and red dots represent the $^{12}C$ and $^{16}O$ mass fractions, respectively, which do not vary significantly in the innermost 0.1~\msun\ of ejecta (see Figure~\ref{explosion}). Progenitors with $M_{ZAMS}=8-12$~\msun\ have comparable mass fractions of carbon and oxygen that are consistent with the observed line intensities in Kes~75.
}
\end{figure}

In the previous sections, we have used a combination of HD simulations, explosion nucleosynthesis models, and shock models to investigate the predicted far-IR line emission from the innermost ejecta of different progenitor models, which we then compared to the line emission observed from the PWN in Kes~75. We chose to test two progenitor models with ZAMS masses of 10.0~\msun\ and 19.8~\msun. These two models qualitatively represent low mass progenitors with compact core structures that result in lower energy explosions, and higher mass progenitors that typically have extended structures and yield higher energy explosions, respectively.
Our HD simulations show that the PWN in Kes~75 has so far swept-up at most 0.1~\msun\ of material, allowing us to compare the nucleosynthetic yields in the innermost ejecta of the two progenitor models with observations.
While the parameters of both models can reproduce the basic properties of the PWN and SNR, as well as the PWN expansion velocity implied by the observed line broadening and roughly consistent with that measured by \citet{reynolds18}, the line intensities observed in Kes~75 favor progenitors that have comparable mass fractions of carbon and oxygen in the ejecta layers currently encountered by the PWN.
The relative mass fractions of carbon and oxygen for the higher mass progenitor cannot reproduce the observed line intensities and ratios due to the relatively low mass fraction of carbon. While the unmixed ejecta model for the 19.8~\msun\ progenitor does not contain carbon in the innermost 0.1~\msun\ of ejecta, the carbon to oxygen mass fraction ratio for the mixed ejecta model is $<< 1$, inconsistent with our observations. Figure~\ref{fig10} shows the relative mass fractions of $^{12}C$ and $^{16}O$ as a function of the progenitor ZAMS mass for the models of \citet{sukhbold16} that assume extensive mixing of the ejecta. The general trend of the increasing fraction of $^{16}O$ and decreasing $^{12}C$ is a consequence of the increasing core size and the $^{12}C(\alpha,\gamma)^{16}O$ reaction during the core helium-burning phase that becomes more dominant in higher mass progenitors. Only progenitors with $M_{ZAMS}= 8-12$~\msun\ have the carbon to oxygen ratios of $\sim$~1, as supported by the observed line emission.

The relative mass fractions of carbon and oxygen for the 10.0~\msun\ progenitor that we modeled are consistent with observations, but these mass fractions can only reproduce the data in the case where the ejecta have not been extensively mixed and where the absolute mass fractions of carbon and oxygen are high enough. Since the lower mass progenitors ($M_{ZAMS}$=8--12~\msun) have more compact cores and lower explosion energies, they should experience less mixing in their ejecta than the higher mass progenitor. The indication of weaker ejecta mixing supported by our analysis is therefore not unexpected. 
The swept-up ejecta mass and SNR age values corresponding to the 10.0~\msun\ progenitor model are 0.05--0.1~\msun\ and 450--620~yr, respectively, consistent with independent estimates from \citet{reynolds18}. While the 10.0~\msun\ progenitor model with unmixed ejecta is likely not a unique solution for Kes~75, our results do show that lower ZAMS mass progenitors with lower explosion energies and comparable abundance ratios of carbon and oxygen in the innermost ejecta are favored.

The SN explosion type is not constrained by our models. While the 10.0~\msun\ progenitor model explored in this work retains its hydrogen envelope and would result in a more typical Type IIp SN explosion, the progenitor of Kes~75 could also have been in a binary system and resulted in a stripped-envelope SN that would have produced a Type Ib/c or a Type IIb SN explosion. In fact, as discussed in Section~\ref{progenitors}, \citet{gelfand14} successfully modeled the dynamical properties of the PWN in Kes~75 using a very low ejecta mass, as may be consistent with a progenitor that lost its envelope to a companion.

\section{Conclusion}\label{conclusion}

We present \textit{Herschel} Space Observatory imaging and spectroscopy of the PWN in the Galactic SNR Kes~75, thought to harbor the youngest known pulsar that underwent magnetar-like outbursts. We detect lines of \ooneline, \othreeline, and \ctwoline\ that show significant broadening with an average corresponding expansion velocity of 730~$\pm$~80~$\rm km\:s^{-1}$. This, in combination with the spatial morphology of the detected line emission, imply that the line emission arises from the innermost SN  ejecta that has been swept-up by the expanding PWN. We use a combination of HD, explosion nucleosynthesis, and shock models to calculate the predicted line intensities and ratios from two different progenitor models; a 10~\msun\ progenitor with a compact core structures and a lower explosion energy, and a 19.8~\msun\ progenitor with a more extended structure and higher explosion energy. While the HD simulations show that both models can reproduce the PWN and SNR properties of Kes~75, the comparison of the observed line emission to their ejecta abundance profiles rules out the higher mass progenitor and favors the 10~\msun\ model in which the post-explosion ejecta is unmixed. The SNR age, PWN expansion velocity, and the ejecta density for this model are 450--620~yr, 630--800~$\rm km\:s^{-1}$, and a few times $\rm 10^{-23}\:g\:cm^{-3}$, respectively. The mass of the ejecta swept-up by the PWN is in the 0.05--0.1~\msun\ range. 
\textit{Herschel} imaging at 70, 100, and 160~\micron\ reveals emission arising from the PWN region that can be explained by a combination of line emission and a continuum arising from a few times $10^{-3}$~\msun\ of dust emitting at a temperature of 33~$\pm$~5~K (45~$\pm$~7~K), assuming silicate (carbon) grains. The dust is most likely SN-formed dust that is being shock-heated by the PWN. The low dust mass is expected given the very low mass of ejecta material so far encountered by the PWN.

While the 10~\msun\ progenitor model for Kes~75 is not unique, this work shows that lower mass progenitors (8--12~\msun) with lower explosion energies and innermost ejecta profiles with comparable abundances of carbon and oxygen are favored. This conclusion provides further evidence that the birth to magnetars may not require unusually energetic SN explosions.

This work demonstrates that observations of young PWNe can be powerful probes of the innermost ejecta produced in SN explosions, and not only provide constraints on the masses of the SN progenitors, but also on the degree of mixing in the explosion. Future observations of young PWNe with the James Webb Space Telescope (JWST) will significantly extend the number of observed emission lines in the near- and mid-IR, providing many more ejecta species (e.g. Ne, Mg, Si, S, Ar, and Fe) for comparison with explosion nucleosynthesis models and important constraints on the SN progenitors and explosion types.

\acknowledgments
We thank Daniel Patnaude and Maxim Lyutikov for the useful discussion and Chris Kolb for his insight on the HD simulation.
PS acknowledges partial support from NASA contract NAS8-03060 and Chandra grant TM6-17002X. BCK acknowledges support from the Basic Science Research Program through the National Research Foundation of Korea (NRF) funded by the Ministry of Science, ICT and Future Planning (2017R1A2A2A05001337).

\bibliographystyle{apj}

\end{document}

%% file: tab1.tex
\begin{deluxetable}{ccl}
\tablecolumns{3} \tablewidth{0pc} \tablecaption{\label{pacsflux}HERSCHEL PACS OBSERVED FLUX DENSITIES}
\tablehead{
\colhead{Wavelength} & \colhead{Flux Density} & \colhead{Ejecta Line Contribution} \\
\colhead{($\micron$)} & \colhead{(Jy)} & \colhead{(Jy)} 
}

\startdata
70.0 & 4.7 $\pm$ 1.5 &  0.31 $\pm$ 0.05 (\ion{O}{1} \& \ion{O}{3}) \\
100 &  5.2  $\pm$ 1.7  &  0.11 $\pm$ 0.02 (\ion{O}{3}) \\
160 &  2.4 $\pm$  1.9 &  0.80 $\pm$ 0.12 (\ion{C}{2})
\enddata
\tablecomments{Background-subtracted infrared flux densities from the PWN in Kes 75 measured from the images shown in Figure~\ref{fig2}. The spectral line contribution to the \textit{Herschel} PACS bands were determined from the spectra shown in Figure~\ref{fig3}.  
}
\end{deluxetable}

%% file: tab2.tex
\begin{deluxetable*}{lcccclcc}
\tablecolumns{7} \tablewidth{0pc} \tablecaption{\label{linetable}Herschel Line Fits}
\tablehead{
\colhead{Line ID} & \colhead{Line Center} & \colhead{Line Flux} & \colhead{FWHM} & \colhead{FWHM} & True FWHM & \colhead{Shift} & \colhead{$ v_{exp}$} \\
\colhead{} & \colhead{(\micron)} & \colhead{($10^{-12} erg\:s^{-1}\:cm^{-2}$)} & \colhead{(\micron)} & \colhead{($km\:s^{-1}$)} & \colhead{($km\:s^{-1}$)} & \colhead{($km\:s^{-1}$)} & \colhead{($km\:s^{-1}$)}
}
\startdata
\sidehead{Source:}
$[$\ion{O}{1}$]$ (63.184) & 63.146 $\pm$ 0.001 & 4.8 $\pm$ 0.7 & 0.269 $\pm$ 0.002 & 1280 $\pm$ 12 & 1270 $\pm$ 100 & -197 & 635 $\pm$ 50 \\
$[$\ion{O}{3}$]$ (88.356) & 88.346 $\pm$ 0.003 & 1.6 $\pm$ 0.2 & 0.459 $\pm$ 0.008 & 1560 $\pm$ 29 & 1550 $\pm$ 120 & -33 & 775 $\pm$ 60 \\
$[$\ion{C}{2}$]$ (157.741) & 157.717 $\pm$ 0.003 & 7.4 $\pm$ 1.1 & 0.832 $\pm$ 0.008 & 1580 $\pm$ 15 & 1570 $\pm$ 50 & -46 & 785 $\pm$ 25\\
\sidehead{Background:}
$[$\ion{O}{1}$]$ (63.184) & 63.206 $\pm$ 0.001 & 2.5 $\pm$ 0.4 & 0.0273 $\pm$ 0.0002 & 129 $\pm$ 10 &  & 107 & \\
$[$\ion{O}{3}$]$ (88.356) & 88.392 $\pm$ 0.001 & 2.1 $\pm$ 0.3 & 0.0489 $\pm$ 0.0002 & 166 $\pm$ 7 &   & 120 & \\
$[$\ion{C}{2}$]$ (157.741) & 157.793 $\pm$ 0.001 & 17.3 $\pm$ 2.6 & 0.1241 $\pm$ 0.0002 & 236 $\pm$ 4 &   & 100 & \\
\enddata
\tablecomments{Best-fit parameters for the spectral fits shown in Figure~\ref{fig3}. The ejecta expansion velocity $v_{exp}$ is taken to be half of the true FWHM and does not include corrections for any geometric parameters.}
\end{deluxetable*}

%% file: tab3.tex
\begin{deluxetable}{lcc}
\tablecolumns{3} \tablewidth{0pc} \tablecaption{\label{dusttab}DUST PROPERTIES}
\tablehead{
\colhead{Composition} & \colhead{Temperature} & \colhead{Mass} \\
\colhead{}  & \colhead{(K)} & \colhead{($\rm M_{\odot}$)}
}

\startdata
Silicates & 33 $\pm$ 5 & 0.044 $\pm$ 0.041   \\
Carbon & 45 $\pm$ 7 &  0.009  $\pm$ 0.006
\enddata
\tablecomments{Dust temperatures and masses for the single-component models shown in Figure~\ref{fig5}.}
\end{deluxetable}

%% file: tab4.tex
\begin{deluxetable}{lcccccc}
\tablecolumns{7} \tablewidth{0pc} \tablecaption{\label{hdinput}HD MODEL INPUT PARAMETERS}
\tablehead{
\colhead{Run} & \colhead{$\rm E_{51}$} & \colhead{$\rm M_{ej}$} & \colhead{$\rm n_{0}$} & \colhead{$\rm \dot{E}_{0,38}$} & \colhead{$\rm n_{psr}$} & \colhead{$\rm \tau_{0}$}\\
\colhead{} & \colhead{($10^{51}$~erg)} & \colhead{(\msun)} & \colhead{($\rm cm^{-3}$)} & \colhead{($\rm erg\: s^{-1}$)} & \colhead{}  & \colhead{(yr)}
}

\startdata

1 &   2.1  &  16.3  &   2.00  &  $1.66\times10^{38}$  & 2.12 &   226 \\
2 & 0.6  &   8.2  &   0.25  &  $1.66\times10^{38}$ & 2.12  &  226
\enddata
\tablecomments{Input parameters to the hydrodynamic models described in Section~\ref{hydro} corresponding to the \citet{bucciantini11} case (Run 1) and the 10~\msun\ ZAMS progenitor case (Run 2). The listed parameters are the explosion energy ($\rm E_{51}$), ejecta mass ($\rm M_{ej}$), the average ambient density ($\rm E_{51}$), and the the pulsar's initial spin-down power ($\rm \dot{E}_{0,38}$), braking index ($\rm n_{psr}$), and spin-down timescale ($\tau_{0}$).}
\end{deluxetable}

%% file: tab5.tex
\begin{deluxetable}{ccccccc}
\tablecolumns{7} \tablewidth{0pc} \tablecaption{\label{progenitors}PROPERTIES OF THE MODELED PROGENITORS}
\tablehead{
\colhead{$\rm M_{ZAMS}$} & \colhead{$\rm M_{NS}$} & \colhead{$\rm M_{ej}$} & \colhead{$\rm M_{He}$} & \colhead{$\rm M_{Fe}$} & \colhead{$\rm M_{Ni}$} & \colhead{$\rm E_{51}$}\\
\colhead{(\msun)} & \colhead{(\msun)} & \colhead{(\msun)} & \colhead{(\msun)} & \colhead{(\msun)} & \colhead{(\msun)} & \colhead{($10^{51}$~erg)}
}

\startdata

10.0  &  1.50 & 8.17  &  2.48  &  1.34  &  0.01  &   0.60   \\
19.8   & 1.59 & 14.3  &  6.12  &  1.45  &   0.07  &   1.96  

\enddata
\tablecomments{Properties of the two different progenitor mass models discussed in Section~\ref{nucleo} \citep{sukhbold16}. The listed parameters are the initial ZAMS mass of the progenitor ($\rm M_{ZAMS}$), the baryonic mass of the neutron star ($\rm M_{NS}$), the ejecta mass ($\rm M_{ej}$), the He-core ($\rm M_{He}$) and Fe-core ($\rm M_{Fe}$) masses of the star at the time of presupernova, the ejected $^{56}$Ni mass ($\rm M_{Ni}$), and explosion energy ($\rm E_{51}$) in units of $10^{51}$~ergs.}
\end{deluxetable}

%% file: tab6.tex
\begin{deluxetable}{lccc}
\tablecolumns{4} \tablewidth{0pc} \tablecaption{\label{abundtab}ELEMENTAL MASS FRACTIONS USED IN THE SHOCK MODELS}
\tablehead{
\colhead{} & \colhead{19.8 $\rm M_{\odot}$} & \colhead{10.0 $\rm M_{\odot}$} & \colhead{10.0 $\rm M_{\odot}$} \\
\colhead{} & \colhead{(mixed)} & \colhead{(mixed)} & \colhead{(unmixed)}
}

\startdata

$^1$H	 & 	\nodata	 & 	1.19E-01 (37.9\%)	 & 	\nodata	 \\ 
$^4$He  	 & 	3.68E-02 (16.1\%)	 & 	7.43E-01 (59.4\%)	 & 	2.09E-03 (0.74\%)	 \\ 
$^{12}$C   & 	2.02E-03 (0.29\%)	 & 	4.57E-02 (1.22\%)	 & 	4.43E-01 (52.5\%)	 \\ 
$^{14}$N 	 & 	\nodata 	 & 	3.64E-03 (0.08\%)	 & 	\nodata	 \\ 
$^{16}$O    	 & 	5.67E-01 (61.9\%)	 & 	4.97E-02 (0.99\%)	 & 	4.30E-01 (38.2\%)	 \\ 
$^{20}$Ne   	 & 	7.62E-03	(0.66\%) & 	7.70E-03 (0.12\%) & 	9.54E-02 (6.79\%)	 \\ 
$^{24}$Mg 	 & 	2.92E-02	(2.12\%) & 	2.32E-03 (0.03\%)	 & 	2.56E-02 (1.52\%)	 \\ 
$^{28}$Si  	 & 	1.70E-01 (10.6\%)	 & 	8.81E-03 (0.10\%)	 & 	2.52E-03 (0.13\%)	 \\ 
$^{32}$S  	 & 	9.27E-02 (5.06\%)	 & 	4.30E-03 (0.04\%)	 & 	3.13E-04 (0.01\%)	 \\ 
$^{36}$Ar   	 & 	1.70E-02 (0.82\%)	 & 	9.19E-04 (0.01\%)	 & 	\nodata	 \\ 
$^{40}$Ca 	 & 	9.71E-03 (0.42\%)	 & 	8.75E-04 (0.01\%)	 & 	\nodata	 \\ 
$^{54}$Fe  	 & 	1.46E-03 (0.05\%)	 & 	1.47E-03 (0.01\%)	 & 	1.57E-03 (0.04\%)	 \\ 
$^{56}$Ni 	 & 	6.54E-02 (2.04\%)	 & 	1.26E-02 (0.07\%)	 & 	5.86E-27 (0.00\%)	
\enddata
\tablecomments{Mass fractions corresponding to the models from \citet{sukhbold16} for two different progenitor ZAMS masses discussed in Section~\ref{nucleo} and shown in Figure~\ref{explosion}. The properties of the progenitors are are listed in Table~\ref{progenitors}. The mass fractions (particle number density percentages) are listed for the extensively mixed ejecta cases and the mildly mixed ejecta case for the 10 $\rm M_{\odot}$ progenitor. The mildly mixed mass fractions are from the zone in which the swept-up ejecta mass is 0.1~\msun\ and are meant to represent the mass fraction range in which the carbon and oxygen fractions are comparable (swept-up ejecta between 0.07 and 0.1 \msun, see Figure~\ref{explosion}).}
\end{deluxetable}